\documentclass[twocolumn,preprintnumbers,amsmath,amssymb]{revtex4}
\usepackage{graphicx}
\usepackage{physics}
\usepackage{color}
\usepackage{gensymb}

\begin{document}

\title{Observation of a Degenerate Fermi Gas Trapped by a Bose-Einstein Condensate}
\author{B.J. DeSalvo}
\author{Krutik Patel}
\author{Jacob Johansen}
\author{Cheng Chin}
\affiliation{James Franck Institute, Enrico Fermi Institute, and Department of Physics, \\ The University of Chicago, Chicago, IL 60637, USA}

\begin{abstract}
We report on the formation of a stable quantum degenerate mixture of fermionic $^6$Li and bosonic $^{133}$Cs in an optical trap by sympathetic cooling near an interspecies Feshbach resonance.   New regimes of the quantum degenerate mixtures are identified. With moderate attractive interspecies interactions, we show that a degenerate Fermi gas of Li can be fully confined in the Cs condensate without external potentials. For stronger attraction where mean-field collapse is expected, no such instability is observed. In this case, we suggest the stability is a result of dynamic equilibrium, where the interspecies three-body loss prevents the collapse. Our picture is supported by a rate equation model, and the crossover between the thermalization rate and the observed inelastic loss rate in the regime where the mean-field collapse is expected to occur.
\end{abstract}

\maketitle

Mixtures of atomic quantum gases serve as an exciting platform to study a rich variety of physics, such as the observation of heteronuclear molecules  \cite{Takekoshi2014, Ni2008,Heo2012,Kraft2006,Roy2016,Park2015,Voigt2009}, Bose and Fermi polarons \cite{Hu2016,Kohstall2011,Schirotzek2009,Scazza2017, Cetina2015}, and superfluid mixtures\cite{Roy2017,FerrierBarbut2014,Yao2016}.
Novel quantum phases have be suggested theoretically \cite{Marchetti2008,Molmer1998,Zaccanti2006,Ospelkaus2006} and probed experimentally \cite{Sengupta2007, Lewenstein2004,Best2009,Gunter2006}.  Intriguing quantum excitations \cite{Santhanam2006, Adhikari2005,Karpiuk2004,Karpiuk2006}, mediated long-range interactions \cite{Bijlsma2000, De2014}, and pairing behavior \cite{Efremov2002,Fratini2010,Gopalakrishnan2015} are proposed based on degenerate Bose-Fermi mixtures. 
 
To date, many experiments exploring degenerate Bose-Fermi mixtures deal with small to moderate mass imbalance \cite{Ospelkaus2006, Zaccanti2006, Taglieber2008, Park2012, Vaidya2015}.  In the case of larger mass imbalance, one expects new phenomena to arise \cite{Fratini2012, Yao2016, Banerjee2007,Gopalakrishnan2015}.  Our choice of light fermionic $^6$Li and heavy bosonic $^{133}$Cs yields the largest mass imbalance among stable alkali atoms. This combination also offers rich interaction properties, which are well characterized \cite{Tung2013,Tung2014,Johansen2016,Ulmanis2015,Repp2013}, and there exist interspecies Feshbach resonances at magnetic fields where  both the Cs Bose-
Einstein condensate (BEC) and Li Fermi gas are stable.  This makes Li-Cs an excellent platform to investigate many-body physics of Bose-Femi mixtures.

\begin{figure}
\includegraphics[clip,trim = 0 1in 0 0.4in,width=3.4 in]{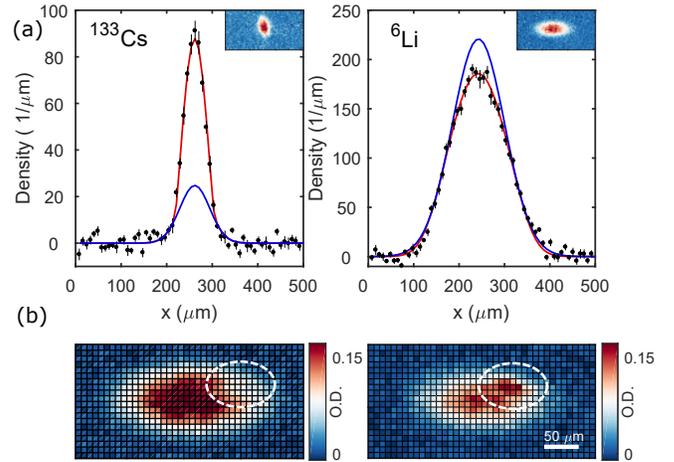}
\caption{Simultaneous quantum degeneracy of $^6$Li and $^{133}$Cs. (a) Left panel: Evidence for Bose-Einstein condensation of Cs is observed as a bimodal distribution of atoms after a time-of-flight expansion.  The blue curve indicates the thermal fraction while the red curve shows a fit to both the thermal and condensate fraction.  Right panel: For Li, a Gaussian fit to the high momentum tail of the time-of-flight images overestimates the density at the center (blue curve).  The full distribution can be fit well using a polylogarithm function (red curve) with $T/T_F = 0.2(1)$ \cite{Ketterle2008}.  In the case of strong interspecies interactions, a modification of the density distribution of the Li is shown in (b).  For repulsive interactions with $a_{BF} = 1000 a_0$ (left panel), a reduction of the Li density is observed at the location of the Cs BEC.  Conversely, an enhancement of the Li density in the same location is clearly visible for attractive interactions with  $a_{BF} = -580$~$a_0$ (right panel).}
\label{Fig1}
\end{figure}

In this work, we explore two novel regimes of dual-degenerate Bose-Fermi mixtures accessed by this combination of atomic species by taking advantage of the tunable interactions afforded by an interspecies Feshbach resonance. First, we find that for small attractive interactions degenerate fermions are found fully trapped and confined by the Cs Bose-Einstein condensate.  Second, at large attractive scattering lengths, we find that the dynamical process of filling the BEC with fermions and enhanced three-body loss can stabilize the mixture against the predicted mean-field collapse \cite{Molmer1998}.

Our experimental procedure to prepare a quantum degenerate Bose-Fermi mixture goes as follows.  After initial laser cooling and optical trapping as described in detail in Ref. \cite{JacobThesis}, we obtain samples of up to $2 \times 10^6$ Li atoms trapped in a deep translatable optical dipole trap, and $2 \times 10^6$ Cs atoms trapped in a separate optical trap.  At this point in the experiment, we have a nearly equal spin mixture of Li atoms in the F=1/2 hyperfine manifold, referred to here as Li$_a$ and Li$_b$, and Cs atoms spin polarized into the  $\ket{F,m_F} = \ket{3,3}$ state, where $F$ is the total angular momentum quantum number and $m_F$ is the magnetic quantum number.  We then ramp the magnetic field to 894.3 G.  This corresponds to a scattering length for Cs of $a_{BB} = 290$ $a_0$ and $a_{\text{Li}_a-\text{Li}_b} = -8~330$ $a_0$ for Li, where $a_0$ is the Bohr radius. This yields efficient evaporative cooling for both Cs and the spin mixture of Li.  Both species are evaporatively cooled over 10 s to approximately 300 nK.  We then remove Li$_b$ with a short resonant light pulse leaving only the absolute ground state of Li and Cs trapped in the optical dipole trap.

These two species exhibit a narrow interspecies Feshbach resonance at 892.64 G \cite{Tung2013, Repp2013, Johansen2016} that can be used to tune the interaction between Li and Cs.  Across the width of this resonance, the Cs-Cs scattering length remains almost constant and varies from $a_{BB} = 220 \sim 280$ $a_0$.  In this range good evaporative cooling efficiency for Cs promises the ability to sympathetically cool Li$_a$ atoms for suitable Li-Cs scattering length. To begin sympathetic cooling, we first ramp the magnetic field to 891 G ($a_{BF} = 20$~$a_0$) and sequentially load both species into a dual-color optical dipole trap comprised of 785 and 1064 nm light \cite{JacobThesis}.  This trap allows for the cancellation of the relative gravitational sag for Li and Cs and ensures good overlap between the two species, even at low temperatures \cite{JacobThesis}.  Depending on whether we want to prepare a mixture with attractive or repulsive interspecies interactions, once the samples are mixed we ramp the magnetic field over 10 ms to either 891.9 or 893.8 G, which yields an interspecies scattering length $a_{BF} = 120$ or $-180$ $a_0$, respectively.  We then perform evaporative cooling for 1.5 s to obtain degenerate samples.  Sympathetic cooling works well for both attractive and repulsive interspecies interactions and both species are able to reach deep quantum degeneracy on either side of resonance.

Detection of quantum degeneracy is performed by analyzing time-of-flight absorption images of both species.  For Cs, after evaporation to the lowest trap depth, we obtain a BEC of $10^4$ atoms with low thermal fraction at a temperature of $T_{Cs} = 20$ nK, as shown in Fig. \ref{Fig1}(a) left panel.  For thermometry of the Li, we first adiabatically ramp the interspecies scattering length over 25 ms to a small value ($|a_{BF}| < 30$~$a_0$) such that the Cs BEC does not influence the Li cloud.  We then release the atoms and image the Li after 1.5 ms expansion.  As shown in the right panel of Fig \ref{Fig1}(a), a Gaussian fit to the high momentum wings of the distribution overestimates the number of atoms at low momentum, indicating the gas is degenerate.  We determine the Fermi temperature of $T_F=480(50)$ nK from the trap parameters, and $T/T_F = 0.2(1)$ from fitting the absorption images using a polylogarithm function \cite{Ketterle2008}.

In the presence of strong Li-Cs interaction, the density distribution of Li can be significantly distorted.  Example images are shown in Fig. \ref{Fig1}(b).  Here, we purposely shift the position of the Cs BEC to the edge of the degenerate Fermi gas to gain visual clarity of the effect. For repulsive interactions (Fig \ref{Fig1}(b) left panel), the Li is repelled from the condensate.  On the other hand, for attractive interactions  (Fig \ref{Fig1}(b) right panel), Li atoms are attracted to the Cs BEC.

According to mean field theory, the potential experienced by one atomic species due to interspecies interactions is given by $ 2 \pi \hbar^2 a_{BF} \Big( \frac{1}{m_{B}} + \frac{1}{m_{F}}\Big) n(\vec{r})$, where  $m_{F(B)}$ is the mass of fermion (boson), $\hbar$ is the reduced Planck's constant, and $n(\vec{r})$ is the density distribution of the other species.  In our case, the density of the Cs BEC is over one order of magnitude larger than the Li degenerate Fermi gas, so the potential experienced by the Li is significantly greater. Given a typical density of a Cs BEC of $n_B = 5 \times 10^{13} $cm$^{-3}$, one obtains that for $a_{BF} = -500$~$a_0$ the trap depth in temperature units felt by Li is 450 nK, which is comparable to the Fermi temperature in the optical dipole trap.  This large depth suggests that Li can be loaded into the BEC even in the absence of another confining potential.

\begin{figure}
\includegraphics[width= 3.4 in]{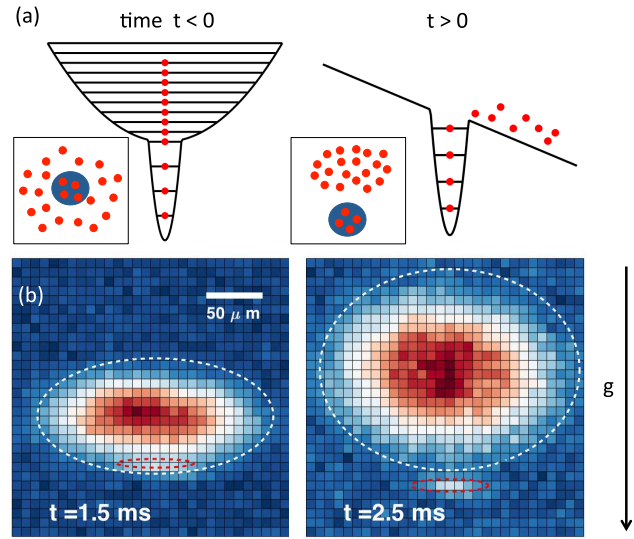}
\caption{Stern-Gerlach separation of Li atoms trapped in a Cs BEC.  As shown schematically in (a), a strong attractive interspecies interaction will confine a small number of Li atoms (red dots) within the Cs BEC (blue circle).  At $t = 0$ we remove the optical trap and apply a magnetic field gradient to separate those trapped from the rest of the sample.  Example images of Li after a short time of flight are shown in (b).  Dashed red circles indicate the position and size of the BEC at the imaging time and the white dashed line indicates the Fermi radius.}
\label{Fig2}
\end{figure}

To investigate this possibility experimentally, we perform a Stern-Gerlach sequence to separate the Li atoms that are trapped by the BEC from those that are not. This is schematically depicted in Fig. \ref{Fig2}a.  We first prepare a degenerate Bose-Fermi mixture in a single beam trap by performing our usual sequence and slowly ramping the intensity of the 785 nm beam to zero.  In this simplified configuration, the overlap of the two species is controlled by the magnetic field gradient, and we obtain trap frequencies of $\omega_{B} = 2\pi \times (6.5, 130, 160)$ Hz for Cs and $\omega_{F}= 2\pi \times(36, 400, 400)$ Hz for Li.  After this preparation, we simultaneously ramp up the magnetic field gradient to 4 G/cm providing a force against gravity and increase the interaction strength to -650$a_0$ over 30 ms.  This deepens the mean-field potential such that a reasonable number of Li can be trapped and shifts the Li up such that the Cs BEC sits on the lower edge of the Li cloud. We then turn off the optical trap, while leaving the magnetic field and gradient on.  The magnetic field gradient is sufficient to over-levitate Li, but not the Cs atoms.  Therefore, Li atoms trapped by Cs BEC will fall downwards.

Results of this experiment are shown in the absorption images of Li after a varying time of flight, see Fig.~\ref{Fig2} (b).  In each image, the white dashed curve shows the maximum extent of the cloud from the calculated Fermi radius and the red dashed curve shows the position and spatial extent of the Cs BEC.  After a time of flight, Li atoms trapped in the Cs BEC are spatially separated from the rest of the sample.  From these images, we also see that the Li atoms are contained in the volume of the Cs BEC and follow its trajectory over the entire time-of-flight.

\begin{figure}
\includegraphics[width= 3.4 in]{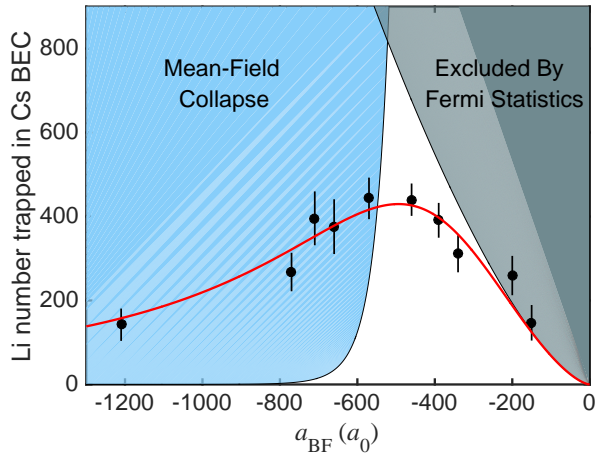}
\caption{Number of Li atoms trapped in Cs BEC.  As a function of interspecies scattering length $a_{BF}$, we measure the number of Li atoms which are trapped in the potential of the BEC after a 1.5 ms Stern-Gerlach separation (black dots).  Within the approximation described in the text, the grey shaded region represents the the region excluded by Fermi statistics and the blue shaded area indicates the region where mean-field collapse is expected .  The red curve shows a fit based on a rate equation model to describe the steady state number of atoms trapped by the Cs BEC (see text).}
\label{Fig3}
\end{figure}

Owing to the Pauli exclusion principle, there is a limited number of Li atoms $N < N_{max}$ that can be trapped within the Cs condensate. Given the approximation that the density of the BEC is not significantly perturbed by Li atoms, the maximum number $N_{max}$ can be found analytically. Within the mean-field and Thomas-Fermi approximations, the density distribution of the condensate is an inverted parabola, yielding a harmonic trap for the fermions with a trap depth $U_0 = \frac{1}{2} (\frac{m_B}{m_F}+1)\frac{|a_{BF}|}{a_{BB}} \mu $ and trapping frequency $\omega_{trap} = [\frac{1}{2}\frac{m_B}{m_F}(\frac{m_B}{m_F}+1) \frac{|a_{BF}|}{a_{BB}} ]^{1/2} \omega_B$, where $\omega_{B}$ is the trapping frequency for Cs and $\mu$ is its chemical potential. 
From Eq. 2, we find that the trap frequency for Li $\omega_F$ is larger than the trap frequency for Cs by a factor of 16 due to mass imbalance alone.   Therefore, across all interspecies scattering lengths of interest, the dynamics of Li in the condensate are much faster than the condensate itself. 

By setting the Fermi energy equal to the trap depth, one finds that the maximum number of Li that can be held by the Cs condensate is
\begin{equation}
N_{max} = \frac{1}{12 \sqrt{2}}\Bigg[\Big(1 + \frac{m_F}{m_B}\Big) \frac{|a_{BF}|}{a_{BB}}\Bigg]^{3/2}\Bigg(\frac{\mu}{\hbar\omega_{B}}\Bigg)^3.
\label{Nmax}
\end{equation}
\noindent This limit is indicated by the grey shaded region in Fig. \ref{Fig3} and is consistent with our measurement for $|a_{BF}|<400$~$a_0$.  For these weak interactions, since the number of Li trapped is close to the maximum, we expect the Li to remain deeply quantum degenerate.  A degenerate Fermi gas confined by a BEC is new type of quantum object that is worthy of future research. 

For a large negative scattering length, mean-field calculations predict another limit on the number of Li atoms that can be trapped by the Cs BEC. When the Li density exceeds a critical value $n_{crit}$, theory predicts a runaway collapse of the sample size due to the loss of mechanical stability. Such critical density for fermions is calculated for a homogenous gas to be \cite{Molmer1998}

\begin{equation}
n_{crit} = \frac{4\pi}{3}\frac{m_B^3m_F^3}{(m_B+m_F)^6}\frac{a_{BB}^3}{a_{BF}^6}.
\end{equation}

In the approximation that the Cs BEC is confined by a harmonic trap and its density distribution is unperturbed by the Li atoms, one can estimate the critical number of Li atoms that can be trapped in a Cs BEC by setting $n_{crit} = n_{p}$, where $n_{p}$ is the peak fermion density in the condensate.  This yields the maximum Li atom number

\begin{equation}
N_{crit} = \frac{16\pi^6\hbar^3}{3\sqrt{2}\omega_B^3} \frac{m_B^{9/2}m_F^6}{(m_B+m_F)^{27/2}}\frac{a_{BB}^{15/2}}{|a_{BF}|^{27/2}}.
\end{equation}

\noindent This limit sets the lower boundary of the shaded blue region in Fig. \ref{Fig3}. Our measurement indicates that for large negative scattering length significantly more Li atoms can be trapped in the Cs condensate than is predicted to be permitted by mean-field calculation.

We propose that the existence of the Li-Cs mixture without mean-field collapse is a result of a dynamical process that includes three-body collisional loss and Fermi statistics. The association of higher particle number with collision loss might seem counterintuitive, but the strong density dependence of three-body collisions is the key to remove atoms preferentially from the high density region and thus prevent the mean-field collapse. 

Since the dynamical time scale for Li is much shorter than for Cs, we suggest the following model for the density of Li atoms trapped in the Cs BEC:
\begin{equation}
\frac{dn}{dt} = A a_{BF}^2 n_F n_B f - B a_{BF}^4 n_{B}^2 n,
\label{DiffEq}
\end{equation}
\noindent where $A$ and $B$ are constants, $n_B$ is the density of the Cs BEC, $n$ is the density of Li confined in the condensate, and $n_F$ is the density of unconfined Li. The first term accounts for elastic collisions which populate the available states in the condensate potential with probability $f\approx 1-N/N_{max}$ given by Fermi statistics.  The second term is the three-body recombination loss due to Li-Cs-Cs collisions \cite{Braaten2006}.  

Due to the small number of Li atoms trapped in the Cs BEC and the large separation in dynamical timescales between the two species, we may further assume that the condensate density profile and Cs number are not greatly disturbed in the time it takes the Li density profile to reach a stationary state. Averaging Eq.~(\ref{DiffEq}) over the extent of the Cs condensate, we obtain $d\bar{n}/dt = A' a_{BF}^2 \bar{n}_B \bar{n}_F f - B' a_{BF}^4 \bar{n}_B^2 \bar{n}$, where $\bar{x}$ is the averaged value of $x$, and $A'$ and $B'$ are constants that incorporate geometric factors from the averaging.

In steady-state $d\bar{n}/dt = 0$, we obtain $N = N_{max}/(1 + C a_{BF}^2 N_{max})$, where $C$ is a constant. Combining this expression with Eq.~(\ref{Nmax}), we fit our data in Fig.~\ref{Fig3} with $C$ being the only fitting parameter. The result yields very good agreement. In the case of no collisional loss $B' = 0$, the fit function reproduces the limit where a deeply degenerate Fermi gas of $N_{max}$ Li atoms is supported by the condensate. 

The observed higher Li number in the Cs condensate above the mean-field calculation can thus be understood as a dynamical process. For large negative scattering length, Li atoms are quickly lost through recombination. Since this process is the highest at the center of the condensate, the loss of Li atoms prevents the runaway density build up at the trap center. Similar loss of Cs atoms also occurs predominately at the trap center. There are however many more Cs than Li within the volume of the condensate, so the loss of Cs is slowed by the limited rate at which its volume is replenished with fermions from the Fermi gas surrounding the BEC. 

\begin{figure}
\includegraphics[clip,trim = 0.0in 0.1in 0.4in 0.4in, width=3.4 in]{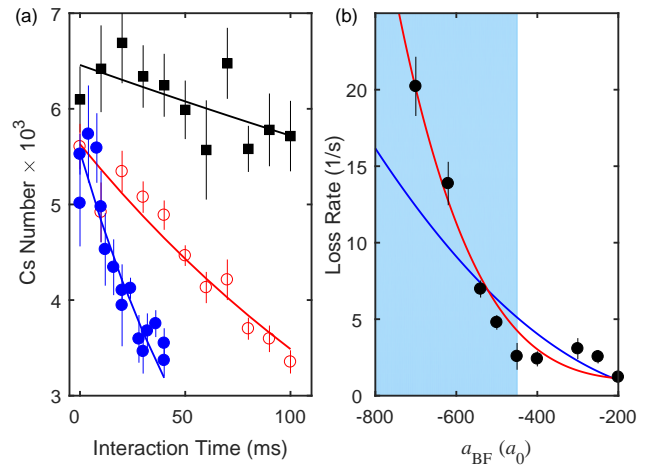}
\caption{Loss dynamics of Cs BEC immersed in Li degenerate Fermi gas. (a)  Atom number in Cs BEC inside of degenerate Fermi gas for $a_{BF}$ = -200 (black), -500 (red), and -620 $a_0$ (blue). The data is well described by a smooth exponential loss (solid lines).  The 1/e loss rates extracted from these fits (b) follow the expected $a_{BF}^4$ scaling from three-body recombination loss (red line).  An estimation of the thermalization rate (see text) is indicated by the solid blue curve. We include a measured constant offset in the loss rate fit to account for Cs-Cs-Cs recombination.  The loss rate exceeds the thermalization rate at $a_{BF} = -520 a_0$, above which the system can no longer reach thermal equilibrium.}
\label{Fig4}
\end{figure}

We seek further experimental support of our picture and, in particular, show that the inelastic process plays a crucial role in the mean-field collapse regime. In Fig. \ref{Fig4}, we study the evolution dynamics of a Cs BEC when it is fully immersed in the Li degenerate Fermi gas. For this experiment, we first create a degenerate Bose-Fermi mixture in the single beam dipole trap, and apply the magnetic field gradient to fully overlap the two species.  After quickly ramping the magnetic field to a desired value, we monitor the number of Cs atoms in the condensate at different interspecies scattering length.  We see that the atom number smoothly decays and the decay rate grows approximately as $a_{BF}^4$ as expected \cite{DIncao2006}. No dramatic drop in atom number associated with the mean-field collapse is observed.

The criterion to validate the mean-field calculation, on the other hand, is the short equilibration time. In our experiment, it is the time it takes for the two species to thermalize. Such thermalization is known to be slow for systems with large mass imbalance. Theoretically, unlike atoms must undergo $\xi = 3(m_B+m_F)^2/4m_Bm_F$ collisions to thermalize \cite{Delannoy2001}. An estimation of the collision rate between degenerate Bose and Fermi gases is in general complicated, and here we employ a model for thermal gases that offers an upper bound on the collision rate given by $\gamma = 4 \pi a_{BF}^2 \bar{v} \int d^3r n_F'(\vec{r}) n_B(\vec{r})$, where $n_F'(\vec{r})$ is the density of fermions available for collisions and the averaged relative velocity of Li and Cs atoms is given approximately by $\bar{v} \approx \sqrt{2E_F/m_F}$ in our experiment. Combining the above results with the fact that the fermion density varies slowly across the BEC, we obtain that the thermalization rate $\Gamma = \frac{N_F' + N_B}{N_F' N_B} \frac{\gamma}{\xi}$ \cite{Mosk2001}, given by
\begin{equation} 
\Gamma \approx 3.74 \frac{m_B m_F^2}{(m_B + m_F)^2} \frac{(N_B + N_F')}{N_F'^{1/3}} \frac{a_{BF}^2 \bar{\omega}_F^2}{\hbar},
\label{Gamma}
\end{equation}
where $\bar{\omega}_F$ is the geometric mean  of the trap frequencies of the fermions, $N_B$ is the number of bosons and $N_F'$ is the the number of fermions available to participate in collisions. We estimate $N_F' \approx g(E_F)k_B T = 3N_F(T/T_F)$  where $g(\epsilon)$ is the single particle density of states, in order to obtain a rough estimation using our thermal framework. This result is shown in Fig \ref{Fig4}(b).  

We see that for large scattering lengths $|a_{BF}| > 520 a_0$, the loss rate exceeds the thermalization rate. This result suggests that for strong attractive interactions, the mixture will quickly deviate from thermal equilibrium. Notably the dominance of inelastic loss is more likely to occur in a Li-Cs system due to its strong mass imbalance. In comparison to the previously studied $^{40}$K-$^{87}$Rb mixture \cite{Ospelkaus2006, Zaccanti2006} where mean-field collapse was observed, the mass imbalance in our system decreases the thermalization rate by a factor of two (from Eq. (\ref{Gamma})), while the three body collision rate is also increased by a factor of two \cite{DIncao2006} for the same trap frequency and atom numbers.  In light of this, our observation of large numbers of Li atoms in the Cs condensate indicates that our system with large attractive Li-Cs interactions reaches a dynamical equilibrium.  Counterintuitively, the three-body loss allows the mixture to survive for a time much longer than would be expected.

In conclusion, we report the first quantum degenerate mixture of Li and Cs and use this system to probe novel regimes of Bose-Fermi mixtures.  For weak attractive interactions, a degenerate Fermi gas with few hundred Li atoms is entirely confined within the condensate.  These atoms can be separated from the untrapped fermions and represent an intruiging quantum object well-suited for future study.  In the case of strong attractive interactions, fast three-body loss prevents the predicted mean-field collapse. The lack of the mean-field collapse in our system allows us to further explore a region of the Bose-Fermi mixture that was previously thought to be inaccessible. 

We thank L.~Feng, C.V.~Parker, K.~Jimenez-Garcia, and S-K.~Tung for help in the early stages of the experiment. We acknowledge funding support from NSF Materials Research Science and Engineering Centers grant DMR-1420709 and NSF grant PHY-1511696. Additional support for B.J.D. is provided by the Grainger Fellowship.


\end{document}